# Mid-infrared intraband transitions in InAs colloidal quantum dots

*Shraman Kumar Saha*[1], *Philippe Guyot-Sionnest*[1]*

[1]Department of Chemistry, and the James Franck Institute, The University of Chicago, 929 E 57th Street, 60637, Chicago, IL, 60653

**Keywords:** InAs, Colloidal quantum dots, Spectroelectrochemistry, Intraband photoluminescence

**Abstract:**

III-V colloidal quantum dots are widely studied for their applications as detectors and emitters from visible to short-wave infrared. They might also be used in the mid-infrared if they can be stably n-doped to access their intraband transitions. Mid-infrared intraband transitions are therefore studied for InAs, InAs/InP, and InAs/ZnSe colloidal quantum dots with an energy gap at 1.4 μm. Using electrochemistry, the quantum dot films show state-resolved mobility, state-resolved electron filling, and intraband absorption in the 3-8 μm range. The InAs/ZnSe films need a more reducing potential than InAs, but the InAs/InP films need a lower reduction potential. As a result, we found that dry films of InAs/InP dots show stable n-doping of the $1S_e$ state, with a steady-state intraband absorption in the 3-5 μm range, and intraband luminescence at 5 μm. low-toxicity, high thermal stability, and stable n-doping, InAs quantum dots become an interesting material for mid-infrared applications.

**Graphics for Table of Contents:**

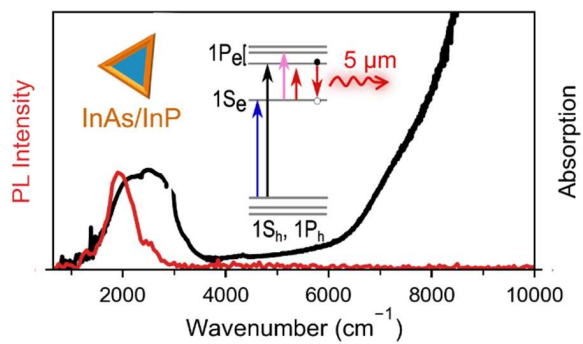

The mid-infrared spectral region (3-5 µm) is widely used for molecular spectroscopy, environmental sensing, machine vision, thermal, and bio-imaging[1]. Conventional mid-IR photon detectors and emitters use single-crystal narrow-bandgap semiconductors (e.g., HgCdTe, InSb) or epitaxial quantum well structures.[2] Colloidal quantum dots (CQDs) offer a promising solution-processed alternative, with tunable band gaps spanning from the visible to the infrared, narrow line width, and high luminescence efficiency. The solution processability of CQD enables cheaper device fabrication compared to traditional crystals. Pb and Hg chalcogenides CQDs already allow photon detection in the short-wave infrared (SWIR)[3] and in the mid-wave infrared (MWIR), respectively.[4,5] In recent years, the toxicity regulations on Pb and Hg have increasingly motivated studies of RoHS-compliant III-V[6–9] and silver chalcogenide[10] CQDs as alternatives in the SWIR. In the mid-infrared (~ 0.25 eV), an alternative to HgTe is to use intraband transitions.[11]

Intraband transitions have been studied in many chalcogenides CQD systems, with demonstrations of infrared detection and emission.[10,12–16] The only prior reports on intraband studies in III-V materials used reduction of InP CQDs by sodium biphenyl,[17,18] but n-doping was unstable in ambient conditions. Although there has been significant improvement in the synthesis and luminescence efficiency of III-V dots, there have been no further reports on intraband studies of III-V colloidal quantum dots.[19] Nevertheless, the high thermal stability and low toxicity of III-V CQDs justify further investigations to determine whether they can be stably n-doped with suitable mid-infrared intraband transitions.

To achieve ambient n-doping in CQDs, the lowest energy conduction band state ($1S_e$) should be below the water reduction potential (-3.7 eV). In principle, the $1S_e$ state should also be below the oxygen reduction potential (-4.9 eV), but kinetic stabilization can be achieved if it is located below -4.0 eV.[20] The bulk conduction band minimum (CBM) sets the lower limit of the $1S_e$ state in CQDs; however, surfaces can shift the energies. InAs should be promising since bulk-InAs has the lowest CBM (-4.9 eV) among other III-V materials[21,22]. The intraband transitions and the n-doping of InAs nanocrystals have not been previously reported but the synthesis has been extensively studied. Shells of materials like InP and/or ZnSe have also been used to enhance their interband photoluminescence and stability.[23–25] This work uses electrochemistry to determine the reduction potentials of films on the quantum dots, the electron mobility, and the intraband transition. It studies InAs CQDs, as cores and with thin shells of ZnSe and InP and it reports the first observation of stable n-doping with III-V quantum dots.

**Results and discussion:**

**Synthesis and characterization of core and core/shell dots.** InAs core, InAs/InP, and InAs/ZnSe core/shell dots are synthesized following a modified procedure previously reported by Ginterseder et. al.[26], Cao et. al[25], and Zhu et. al[23], respectively. The synthetic scheme is shown in Scheme 01, and the detailed synthetic procedures are reported in the method section. Except for InAs/ZnSe, the dots are prone to oxidation, and all post-synthetic processing is done under inert conditions unless otherwise mentioned. Surface oxides can be reduced by the injection of electrons, so that all oxygen-containing reagents (e.g., fatty acids as ligands) are avoided while synthesizing the cores and during shell growth. Figure 1a shows the steady state absorption spectra of InAs cores grown at 320 °C. The absorption shoulder is around ~7200 cm$^{-1}$ or 1.4 μm. The optical absorption spectra do not show major effects of these thin shells, consistent with both InP and ZnSe confining the electronic states of the InAs cores. The InP shell is grown at 240 °C. It does not broaden the absorption spectrum, and it leads to a small red shift. This indicates that the size distribution is preserved and that there is no alloying, which would induce a blue shift. The ZnSe shell is grown at a higher temperature of 300 °C. It leads to a lesser shift but also results in more broadening. The absorption spectra of InAs/InP core/shell dots (Figure 1a) show two shoulders, at ~7200 cm$^{-1}$ and ~9500 cm$^{-1}$. A peak fitting (Figure S01, and Table S1) was performed on the absorption spectra of the InAs/InP core/shell with three Gaussians: The first (1S) is centered around ~ 7250 cm$^{-1}$, the second (1P) is centered around ~ 9200 cm-1, and the third accounts for higher order transitions. Assuming that the heavy-hole dispersion is negligible, the intraband transition ($1S_e \rightarrow 1P_e$) energy should be similar to the difference between 1P and 1S transitions, i.e., ~2000 cm$^{-1}$ (5 μm), within the desired mid-infrared region.

The powdered-XRD spectra (Figure 1b) confirm that cores and core/shells are zinc blend. The Debye-Scherrer analysis of the XRD peak around 25.5°, assuming a spherical shape, shows a size increase after shell growth. The increase is ~1.8 ML of InP using a thickness of 0.34 nm for an InP monolayer and ~2 ML of ZnSe using a thickness of 0.33 nm for a ZnSe monolayer. (See Discussion 02 in SI)

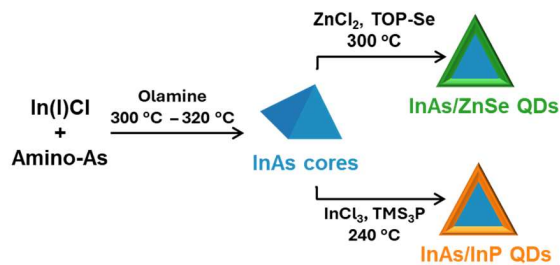

**Scheme 01:** Schematic representation of the synthesis of and shell growth on InAs cores.

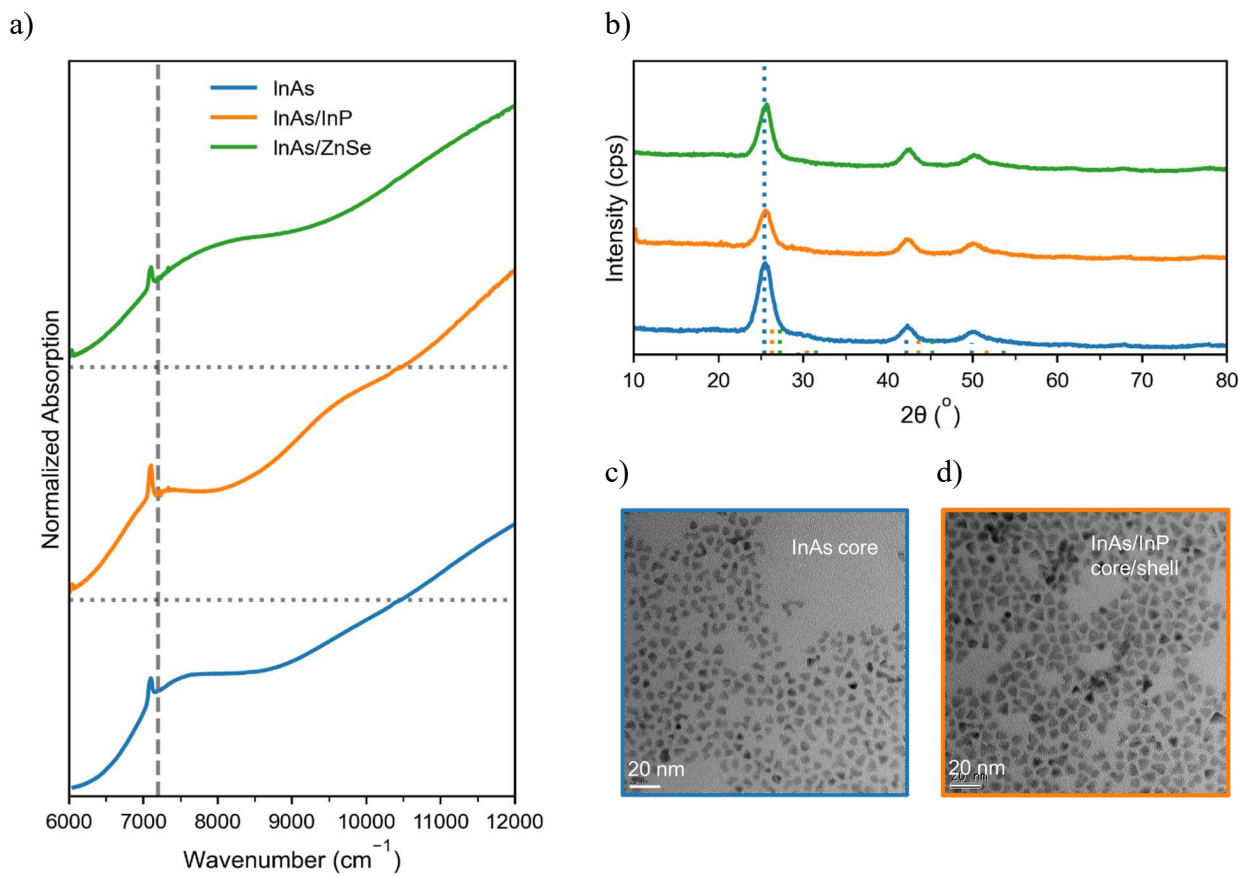

**Figure 01:** a) Normalized steady state absorption spectra of washed InAs core (blue), InAs/InP core/shell (orange), and InAs/ZnSe core/shell (green) dots. The small feature at ~ 7050 cm$^{-1}$ is from residual ethanol in the sample. The dotted line indicates the zero line for the respective spectra. b) Powder XRD pattern of InAs core, InAs/InP core/shell, and InAs/ZnSe core/shell dots, the vertical line represents the corresponding position and intensities of X-ray diffraction of bulk InAs (blue), bulk InP (orange), and bulk ZnSe (green). TEM images of c) InAs core and d) InAs/InP core-shell dots.

Transmission electron microscopy (TEM) shows that the shapes of the dots are actually angular. Images of InAs cores and InAs/InP core/shell are shown in Figures 1c and 1d. Images of InAs/ZnSe core/shell are shown in Figure S05. Cores and core/shells dots are crystalline (Figure S03c, S04c, and S05c) with fair monodispersivity (Figure S03d, S04d, and S05d),s and there is no evidence of secondary nucleation upon shelling. Using a distance from the middle of an edge to a point, the mean size of the InAs core only, InAs/InP core/shell, and InAs/ZnSe core/shell dots are 7.5 +/- 0.8 nm, 8.4 +/- 1 nm, and 9.5 +/- 1 nm, respectively. The size from TEM indicates ~1.5 ML InP shell, and ~ 3 ML ZnSe shell, roughly consistent with Debye-Scherrer analysis. Histograms of the sizes are obtained from TEM images (See Discussion 03, Table S3 in SI). The sizes of individual dots are larger than those obtained from the Debye-Scherrer equation, maybe because the shapes are not spherical.[27] The TEM images of InAs/ZnSe (Figure S05) show a more irregular shape, which may be consistent with Ostwald ripening at the higher temperature used.

**Electrochemical state filling in the dots.** Prior measurements of carrier mobility with InAs CQD films used solid-state field effect transistors (FET).[28,29] They did not yet resolve state filling features, nor was transport correlated with spectroscopy. A consequence is that the states responsible for transport cannot, a priori, be assigned to the quantum dots states instead of other surface or defect states. Electrochemical gating can more readily display state-filling because it can inject many more charges per dots.[30] The InAs cores are drop-cast and cross-linked with ethylene diamine on an interdigitated electrode (IDE) substrate. The IDE substrate has a finger spacing of 50 µm, a finger width of 12 µm, and a total length of 30 x 0.2 = 6 cm. The Ag-wire reference electrode is calibrated after the experiment by adding Ferrocene/Ferrocenium. The electrochemical potentials ($E_{app}$) are reported with respect to NHE and to the vacuum level (Discussion S4). The Faradaic current from impurities is minimized by cooling the electrochemical cell in a dry ice-ethanol bath at ~ -72 °C. The cyclic voltammetry is performed from 0 V vs Ag wire (-4.5 eV) to -0.7 V vs Ag wire (-3.8 eV) with a slow scan rate of 2 mV/s to lower the capacitive currents. The anhydrous electrolyte is 0.1 M TBAClO$_4$ solution in acetonitrile.

The bipotentiostat records the current from both electrodes, $I_1$ and $I_2$ from which the charging current is $I_c = I_1 + I_2$ and the conduction current is $I_s = (I_2 - I_1)/2$.[30] Figure 2a shows the charging current and the conduction current of InAs core-only dots. At the less reducing potential, the conduction current is negligible compared to the charging current, consistent with the absence of carriers. At more reducing potentials, the conduction current dominates because more electrons are injected. Both the charging and conduction currents overlap well for each scan, showing a good reversibility of the electrochemical doping. The hysteresis of the charging current reflects the charging and discharging of the film. The conduction current also shows some hysteresis that diminishes when it dominates the capacitive current. This is assigned to imperfect symmetry of the electrodes.

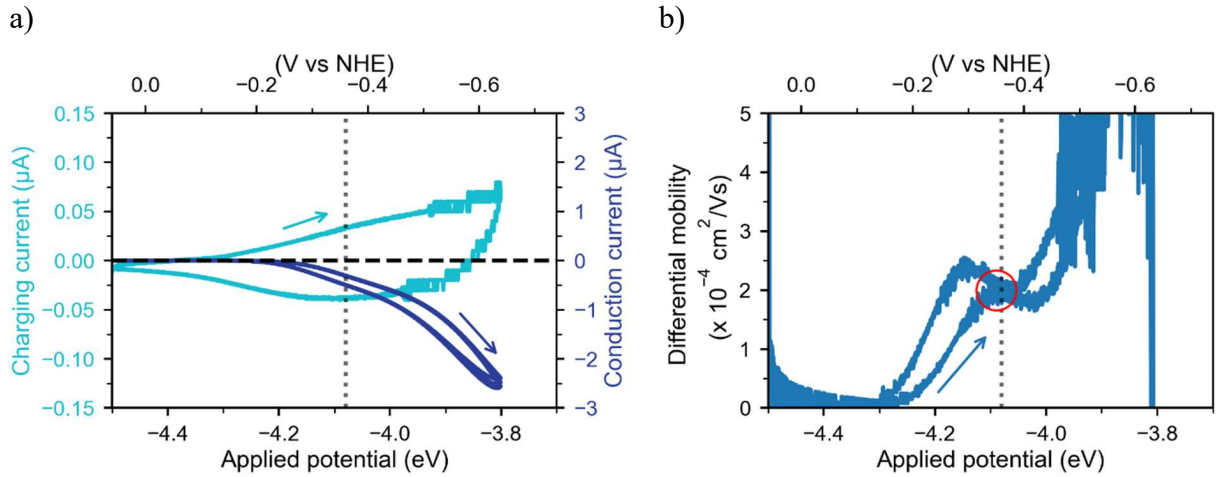

**Figure 2:** a) Cyclic voltammogram of InAs cores drop-cast on the IDE substrate, showing the charging current (cyan) and the conduction current (blue). The bias between the electrodes is -0.1 V. b) Differential mobility of the InAs cores. The direction of the arrows indicates the scan direction.

The conductance of the film is given by $G = I_s/V_b$ where $V_b$ is the fixed bias between the electrodes (-0.1V in Figure 2). The differential mobility is shown in Figure 2b and it is calculated using:

$$\mu = \frac{\frac{dG}{dV_G}(2d_1 + d_2)d_2}{\frac{d\left(\int_{t=0}^{t} I_c \, dt\right)}{dV_G}}$$

(1)

where $d_1$ is the width of the metal fingers (12 µm) and $d_2$ is the gap between the metal fingers (50 µm), $V_G$ is the applied electrochemical bias. The mobility starts in the noise, increases from -4.3 eV onwards, peaks at ~3 x $10^{-4}$ cm²/Vs around -4.15 eV, and decreases again. The peak is assigned to the initial filling of $1S_e$, at about 0.5 e⁻ per dot. The second rise is assigned to filling the $1P_e$ states. The mobility is relatively small for dots of ~7 nm size, reflecting a low inter-dot electronic coupling. Much higher mobilities have been obtained following aggressive ligand exchange and thermal annealing but this mobility is sufficient to perform spectroelectrochemistry on films.[28] Indeed, the electron diffusion time across a film of thickness $l \sim 100$ nm is $\Delta t = \frac{el^2}{k_B T \mu} \sim 13\ \mu s$. Therefore, the charging speed is limited by the ion mobility in the film and not by the electron mobility.

**Spectroelectrochemistry.** Films were drop-casted on a gold-coated stainless-steel working electrode of 8 mm diameter, and with the same ligand exchange with 1% ethylenediamine/IPA solution. The window is a KBr disk. The spectroelectrochemical cell is filled in the glovebox and sealed before being taken out for measurements. The infrared absorption is measured with the cell in the nitrogen-filled compartment of the FTIR spectrophotometer. Spectra are taken at room temperature in reflection mode. Before the measurements, the electrode is gently pressed against the KBr window to minimize the absorption of the infrared light by the electrolyte. As the electrochemical potential is applied, the spectra are recorded relative to a background spectrum. This background spectrum is taken at a potential where the dots are mostly not doped, 0V vs Ag for the InAs cores and +0.4V vs Ag for the InAs/InP, and InAs/ZnSe core/shell dots. The change of optical density ($\Delta OD$) with applying potentials for InAs cores, InAs/InP core/shell, and InAs/ZnSe core/shell are shown in Figures 3a, 3b, and 3c, respectively.

Upon a reducing potential, an interband bleach feature, marked as $e_1$, appears at ~7200cm$^{-1}$ (~1.4 µm) (Figure 3a). The bleach is assigned to electron transfer to the lowest energy electronic state of the conduction band, labeled $1S_e$. It is a single Gaussian, indicating that a relatively narrow set of hole states has allowed transitions to $1S_e$. Synchronous with $e_1$, there is a broader induced absorption in the mid and long-wave infrared, between 1200 cm$^{-1}$ and 3300 cm$^{-1}$ (8 µm to 3 µm). This absorption has some structure, with at least two clear features, labeled $i_1$ and $i_2$. The relative overall oscillator strength of the induced absorption and bleach is similar, and they both increase with more reducing potential before saturating. This suggests that $i_1$ and $i_2$ are electron transitions from $1S_e$ to higher states, consistent with the peak fitting of the absorption spectra of InAs/InP core/shell dots. Excursions to positive potential degraded the films irreversibly. This indicates that InAs core dots could not be p-doped within the experimental conditions.

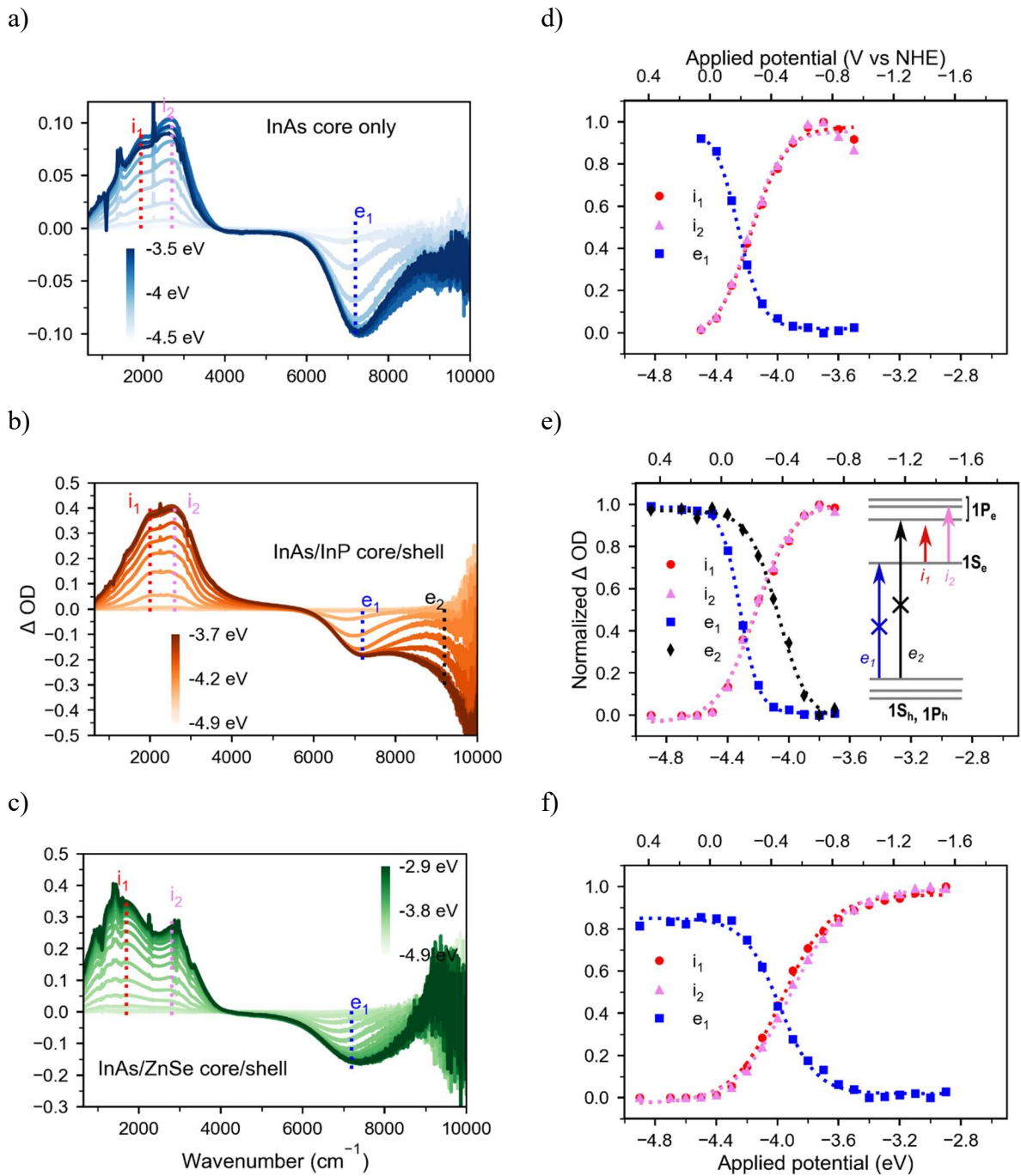

**Figure 3:** (a, b, c) Intraband absorption and interband bleaching of InAs, InAs/InP, and InAs/ZnSe dots, respectively. The red and violet dashed lines ($i_1$ and $i_2$) follow the induced intraband absorption features, and the blue dashed line ($e_1$) follows the bleached interband transition. (d, e, f) Normalized change in optical density as a function of applied potential for core and core/shell dots, respectively; the dashed lines are sigmoidal fits. The inset image in 2e is a schematic representation of electrochemically induced optical transitions. $e_1$ (blue) and $e_2$ (black) arrows denote interband absorptions, which bleach upon state filling, while $i_1$ (red) and $i_2$ (purple) arrows indicate induced intraband absorption.

Figure 3d, 3e, and 3f show the normalized induced absorptions (at points $i_1$ and $i_2$) and bleach ($e_1$ and $e_2$) of InAs cores, InAs/InP, and InAs/ZnSe core/shell dots, respectively, as a function of applied reducing potential, as a Nernst plot. The Nernst plot for $i_1$ and $i_2$ (Figures 3d, 3e, and 3f) overlaps for all three different systems, ruling out the possibility that $i_1$ and $i_2$ are due to different sizes of dots. Therefore, $i_1$ and $i_2$ are assigned to at least two states to which the $1S_e$ electrons can be excited. In a 'particle in a sphere' model, the selection rules allow transition of electrons from $1S_e$ to $1P_e$, while transitions to other states are weak or forbidden. $i_1$ and $i_2$ may originate from splitting of the $1P_e$ state, from spin orbit coupling or shape effects [31,32]. This should be determined by future electronic calculations of InAs tetrahedral cores.

The Nernst plots (Figures 3d, 3e, and 3f) show that the normalized $\Delta OD$ of InAs and InAs/InP is reversible between -4.8 eV and -3.5 eV and starts to decrease around ~ -3.5 eV, whereas InAs/ZnSe sustains more reducing potentials around ~-3 eV. This indicates that InAs and InAs/InP are less stable than InAs/ZnSe core/shell dots against irreversible reduction. Besides transferring to CQDs states, injected electrons can reduce ligand and surface cation, and participate in irreversible chemical reactions.[33] The electrochemical stability of reduced semiconductor CQDs can be discussed following Gerischer, where the electrons can degrade the CQDs.[34] One of the mechanisms is the surface cation reduction reaction as $(MX)_n + ze^- \rightarrow M^o + X^{z-} + (MX)_{n-1}$. The reducing potential of the surface cation reduction is in the order ZnSe > InP ≥ InAs, i.e., ZnSe has the most reducing potential (See discussion 05, Table S4 in SI). This may explain why the ZnSe shell helps stabilize InAs, but there can be other possible pathways for electron capture.

Figure 4a shows the normalized spectra of the InAs and InAs/InP dots films, at their respective maximum applied reducing potential. The narrowest intraband spectra and interband bleach are observed at a weakly reducing potential for InAs/InP, and this is assigned to the preferential reduction of the larger particles (Figure 4b). InAs/ZnSe has the broadest intraband spectra consistent with the more irregular shape seen in the TEM. The intraband absorption spectra also extends to quite low energy, below 650 cm$^{-1}$, and this low energy tail might be avoided in future work, with more compact shapes and a narrower size distribution.

A higher energy interband bleach, labeled $e_2$, is also observed (Figure 3b and 3e) in the spectroelectrochemical spectra of InAs/InP core/shell dots, and less so for InAs cores (Figure 3a).

Figure 4c shows differences between consecutive spectroelectrochemical spectra for InAs/InP. Around -4.0 eV, Figure 4c, there is no further bleach of $e_1$, and all the bleach is in $e_2$. This can be explained by a filled $1S_e$ state and the filling of the lower energy $1P_e$ state (corresponds to $i_1$ transition). Indeed, the intraband absorption gains strength at $i_2$ and on its blue side, and this is assigned to transitions from the $1P_e$ state to higher transitions such as $1D_e$ or $2S_e$. These signatures of $1P_e$ state filling are not seen for InAs/ZnSe dots (Figure S07b), while the bleaching of $e_2$ is much weaker for InAs (Figure S07a).

The Nernst plots (Figure 3d, 3e, and 3f) are fitted to a modified Nernst equation using the following relation:

$$\Delta OD \propto \frac{1}{1 + e^{\left(\frac{E_{app}-E_o}{w}\right)}}$$

(2)

Where $E_{app}$ is the applied potential, $E_o$ is a fitted mean reduction potential which will be taken as the absolute energy position of the concerned state referred to mean potential hereafter, and $w$ is a fitting parameter that is allowed to deviate from the Nernst value of $\frac{k_BT}{e} = 25$ mV for a one electron reduction step at room temperature.[35] $w$ can then include charging energy, electron repulsion, and inhomogeneity of the redox potential. The fit parameters are reported in Table S5. Figure 3d summarizes the mean reduction potential, obtained from Nernst fitting, for $i_1$, $i_2$, and $e_1$ transitions for three sets of dots. The mean potential for $i_1$ and $i_2$ transitions is the highest for the InAs/ZnSe dots (~ -3.95 eV), the lowest for the InAs/InP dots (~ -4.3 eV), and in between for InAs dots (~ -4.2 eV). This 1Se peak obtained from electrochemical gating (~ -4.1 eV, Figure 2b) is slightly more reducing than the mean potential of the $1S_e$ state (~ -4.3 eV, Figure 4d, and Table S5) obtained from the Nernst plot, and this could be due to resistance across the rather large electrode gap of 50 μm.

The black dashed line in Figure 4d is a simple two-band **k·p** prediction of the mean potential for the InAs cores band edge absorption near ~7200 cm$^{-1}$ or 1.4 μm (see discussion 08 in SI). The predicted mean potential is closest to the InAs/InP data, and furthest from InAs/ZnSe. Absolute potential can easily deviate from bulk values by strain, composition, and surface properties, and atomistic theoretical studies are needed. However, the data show that InP shell growth results in a relatively low mean reduction potential.

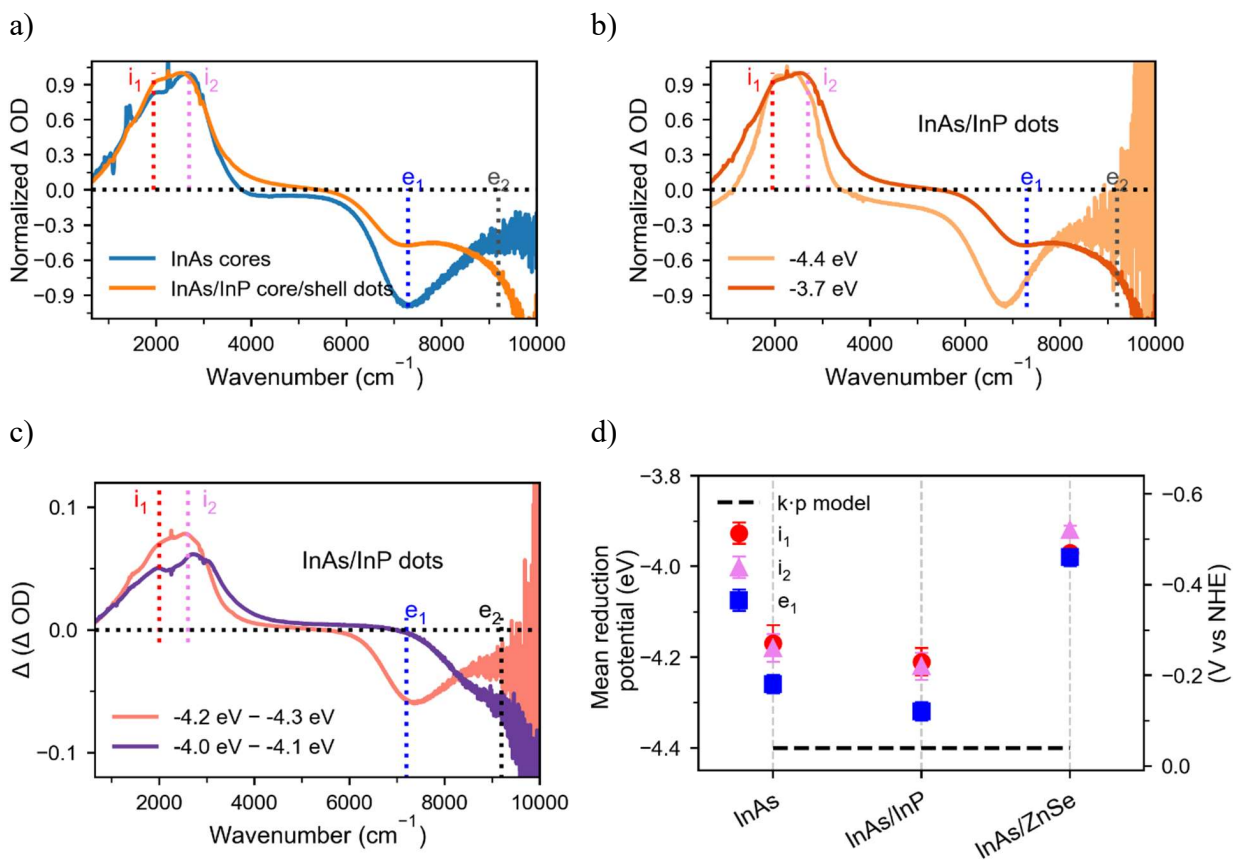

**Figure 4:** a) Normalized spectroelectrochemical spectra of InAs core and InAs/InP core/shell dots at their respective maximum reducing potential. The InAs/InP spectrum exhibits more bleaching at $e_2$, consistent with filling of the 1Pe state. b) Normalized spectroelectrochemical spectra of InAs/InP dots at -4.4 eV (lower reducing potential), and -3.7 eV (higher reducing potential). The intraband spectra at lower reducing potential are narrower than at higher reducing potential. c) $\Delta OD$ at two consecutive reducing potential for InAs/InP. The higher reducing potential shows more bleaching near $e_2$ and a blue shift of the intraband transition compared to lower reducing potential. d) Mean reduction potential for various InAs core-based dots obtained from the Nernst fits. The dashed line represents the calculated value from the **k·p** model.

**Infrared absorption and photoluminescence measurements.** We therefore looked for the possibility that InAs/InP may be stably n-doped. The steady state absorption of a film is measured using the Attenuated Total Reflection (ATR) FTIR method, where the dots are drop-cast on a ZnSe plate without any crosslinkers. The photoluminescence (PL) is measured using films on an aluminum substrate, and a 150mW modulated diode laser at 808 nm is used for excitation and lock-in detection. The film prepared for PL measurements was washed with isopropyl alcohol to remove excess ligands. No interband or intraband PL is observed for InAs core dots. With InP/ZnSe, only interband PL is observed (Figure S09), consistent with prior studies[23,36]. On the other hand, for InAs/InP, Figure 5 shows the steady-state absorption and the PL spectra. The emission is around ~ 2000 cm$^{-1}$, at the same position as the $i_1$ transition. The absence of PL at the $i_2$ transition is expected given the proposed 1Pe state splitting and Kasha's rule.[37] The film can then be exposed to ambient air. While the outer surface likely oxidizes, much of the intraband PL is preserved (Figure S10). The PL is quite weak, at best about 1/10$^{th}$ of the signal observed in the same setup with the interband emission of undoped mid-IR HgTe CQD solutions, but this may be because only a small fraction of the InAs/InP dots are n-doped.

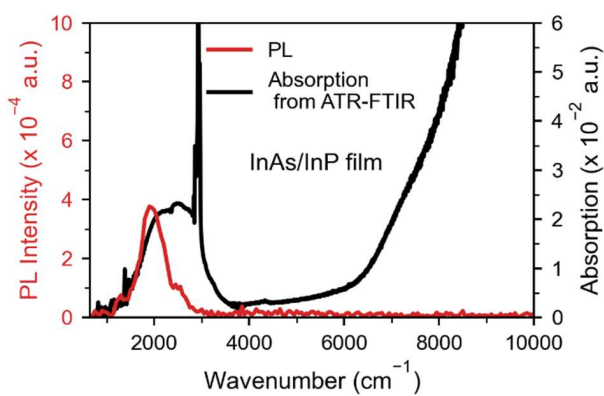

**Figure 5:** The steady-state intraband photoluminescence (PL) spectrum (red) of InAs/InP core/shell quantum dots drop-cast on an aluminum film and protected from air by a CaF$_2$ window. The PL is compared with the steady state intraband absorption spectrum of the same dots (black).

**Conclusion**

In conclusion, films of InAs core-based dots with a band edge absorption around 1.4 μm are studied by electrochemistry for the first time. Using solid-state ligand exchange with ethylenediamine, the films of InAs cores showed state-resolved electron mobility in the 10$^{-4}$ cm$^2$/Vs range. This is sufficient to perform infrared spectroelectrochemistry and to observe the changes of occupation of the quantum dot states. Upon reduction, InAs cores show a strong reversible induced intraband absorption in the 8-3 μm range along with the bleach of the first interband exciton at 1.4 μm. In contrast, on the oxidation side, no spectral changes are observed before film degradation. Spectroelectrochemistry therefore confirms that electrons can be injected into the $1S_e$ state of the InAs dots while no holes can be injected. Thin shells of 1-3ML of InP and ZnSe were then studied to explore their surface effects on the electrochemical properties. All films can be n-doped by electrochemistry. The ZnSe shell increases the reduction potential of the $1S_e$ state but the $1P_e$ state could not be charged even though the films are stable until a quite reducing potential. In contrast, the InP shell lowers the reduction potential and allows more facile electron occupation of the $1S_e$ state and even of the $1P_e$ state. This confirms that the surface of the CQDs plays a critical role in the electrochemistry and the doping tendencies. A particularly significant result is that the films of the InAs/InP CQDs show stable n-doping in ambient air, with steady state absorption and photoluminescence around 5 µm. Larger InAs CQDs will have lower reduction potentials and will be even easier to n-dope. Therefore, this works expands the scope of InAs CQDs, since they can now be studied for detection and emission at energies below the bulk gap of InAs, while they combine a lower toxicity or a higher thermal stability than all currently explored chalcogenides.

**Materials and Methods:**

**Chemicals:**

Indium (III) chloride (anhydrous, 99%-metals basis, Thermo-Fischer), Indium (I) chloride (anhydrous, 99.99%-In, Puratrem), Tris(dimethylamino)arsine (99%, Strem chemicals), Selenium powder (~ 100 mesh, 99.99% trace metal basis, Sigma-Aldrich), Zinc chloride (anhydrous, >= 98%, Sigma-Aldrich) Tris(trimethylsilyl)phosphine (95%, Sigma-Aldrich), Oleylamine (70%, technical grade, Sigma-Aldrich), Octadecene (90%, technical grade, Sigma-Aldrich), Trioctylphosphine (97%, Sigma-Aldrich), Tetrachloroethylene (>= 99%, anhydrous, Sigma-Aldrich), Ethanol (200 proof, Sigma-ALdrich), 1,2-ethanediamine (>= 99%, Sigma-Aldrich), Isopropanol (99.5%, anhydrous, Sigma-Aldrich), Acetonitrile (99.8%, anhydrous, Sigma-Aldrich), Toluene (99.8%, anhydrous, Sigma-Aldrich), Tetrabutylammonium perchlorate (>= 99.0%, Sigma-Aldrich), Ferrocene (Sigma-Aldrich). Oleylamine and Octadecene were degassed at 130ºC for 4 hours before their use. All the other chemicals were used without any further modifications.

**Amino-Arsine stock solution:**

Arsenic precursor stock solution was prepared by following the method of Srivastava et. al[38]. 72 µL of Tris(dimethylamino)arsine (~ 0.04 mmol) was dissolved in 1 mL degassed oleylamine. The mixture was heated at 50 °C until bubbles stopped evolving. The concentration of the resulting solution is ~0.04M.

**TMS$_3$P stock solution:**

1 g of Tris(trimethylsilyl)phosphine was diluted with 1 mL of degassed octadecene. The concentration of resulting solutions is ~0.85M.

**InCl$_3$-TOP stock solution:**

290 mg of InCl$_3$ was dissolved in 10 mL of Trioctylphosphine by stirring the solution at 120 °C. The concentration of the resulting solution is 0.13M.

**InAs core synthesis:**

The InAs core synthesis was performed using a modified procedure of Ginterseder et. al[26]. 0.5 mmol of In(I)Cl and 7.2 mL degassed oleylamine were loaded in a 50 mL three neck round round-bottom flask inside a nitrogen glovebox. The flask was transferred to the Schlenk line and purged with Argon gas. The temperature of the mixture was set to 320 °C. When the temperature reached 300 °C, 0.4 mL of 0.04 M As precursor solution was injected into the mixture swiftly. The reaction was allowed to continue for another 40 minutes. The system took 20 minutes to increase the temperature from 300 °C to 320 °C, and another 20 minutes, the reaction continued to stir at 320 °C. After 40 minutes, the reaction mixture was allowed to cool and transferred to the glove box. The reaction mixture was diluted with another 7.2 mL of Toluene and stored in a 20 mL scintillation vial at room temperature. The diluted reaction mixture was referred to as a crude solution hereafter.

**InAs core washing:**

0.5 mL of ethanol was mixed with 0.5 mL of crude solution and centrifuged at 9000 rpm for 1 minute to precipitate all the nanocrystals. The supernatant was clear and discarded. The precipitate was redispersed in 0.2 mL TCE and centrifuged again to remove all the unwanted precipitate. Another 0.2 mL of ethanol was added to the supernatant and centrifuged to precipitate the dots. The dots were dispersed in 0.1 mL TCE for further characterization.

For core/shell synthesis, 5 mL of crude solution was washed following the above-mentioned solvent to anti-solvent ratio and redispersed in 1 mL Toluene instead of TCE.

**InAs/InP core/shell synthesis:**

The InAs/InP core/shell synthesis was performed using a modified procedure of Cao et. al[25]. 3.6mL of TOP was loaded in a 50 mL three-neck flask and purged with Argon gas. Inside the glovebox, 260 µL of 0.13M $InCl_3$-TOP solution and 16 µL of 0.85M $TMS_3P$ in ODE solution were mixed at room temperature. This is our InP precursor solution. The temperature of the TOP was increased to 240 °C. The 1 mL InAs-toluene solution was injected into the TOP at 240 °C swiftly, followed by the injection of InP precursor. The reaction was allowed to stir at 240oC for 10 minutes. After 10 minutes, the reaction was cooled down to room temperature and transferred to the glovebox. The reaction mixture was diluted with another 3.6 mL of Toluene.

The InAs/InP core/shell dots were washed with the same solvent to anti-solvent ratio as InAs cores.

**InAs/ZnSe core/shell synthesis:**

The InAs/InP core/shell synthesis was performed using a modified procedure of Zhu et. al.[23]. 3.6mL of Olamine and 81 mg $ZnCl_2$ were loaded in a 50 mL three-neck flask and purged with Argon gas. Inside the glovebox 80 mg of powdered Se was dissolved in TOP solution at 110 °C to prepare a 1M TOP-Se solution. The 1 mL InAs-toluene solution and 150 µL of 1M TOP-Se were injected into the TOP at room temperature. The temperature of the mixture was increased to 300 °C. The reaction was allowed to stir at 300 °C for 15 minutes. After 15 minutes, the reaction was cooled down to room temperature and transferred to the glovebox. The reaction mixture was diluted with another 3.6 mL of Toluene.

The InAs/ZnSe core/shell dots were washed with the same solvent to anti-solvent ratio as InAs cores.

**UV-Vis-NIR absorption spectroscopy:**

The nanocrystals dispersed in TCE were used for absorption measurements. The absorption spectra of the solutions were collected using a Shimadzu UV-3600 Plus UV-VIS-NIR spectrophotometer.

**ATR-FTIR absorption spectroscopy:**

Dots were washed and dispersed in TCE as mentioned previously. This TCE solution is drop-casted on a ZnSe ATR plate. The plate was kept in an air-tight cell, and the ATR-FTIR absorption spectra were measured using the Thermo Nicolet iS50 Advanced FT-IR machine.

**Powder X-Ray diffraction measurement:**

The diffraction patterns were obtained using a Bruker D8 diffractometer with a Cu Kα X-ray source operating at 40 KV and 20 mA and a Vantec 2000 area detector.

**Transmission Electron Microscopy measurements:**

The images were obtained using a 300 kV FEI Tecnai F30 microscope. Samples for TEM were prepared by depositing one droplet of diluted nanocrystal solution in TCE onto 400 mesh Formvar/Carbon supported Copper grids.

**Spectroelectrochemistry:**

The nanocrystal solutions in TCE were drop-cast and then cross-linked using a 1% ethylenediamine/IPA solution on a polished gold-coated stainless steel surface. An Ag wire pseudoreference electrode and a Pt counter electrode were placed in the electrochemical cell near the working electrode. The spectroelectrochemical cell was assembled and filled with 0.1 M tetrabutylammonium perchlorate (TBAClO$_4$) in acetonitrile. The sample electrode was then pressed gently against the KBr window to minimize the infrared absorption from the electrolyte. Then the cell was placed in an FTIR instrument (Thermo Nicolet iS50 Advanced FT-IR machine). After the sample had been flushed with N$_2$ for 10 min, infrared spectra were measured in reflectance mode. We first set the potential to the open circuit potential, or any desired value to take the background, and the difference spectra were taken across a wide potential range. For InAs cores the potential window was 0 V to -1V vs Ag wire, for InAs/InP core/shell it was +0.4 V to -0.8V vs Ag wire, and for InAs/ZnSe it was +0.4 V to -1.6 V vs Ag wire. With potential more negative than −0.8 V vs Ag wire (for InAs core only) and -0.5 V vs Ag wire (for InAs/InP core/shell), the films became permanently altered, unable to return to their initial state.

After each spectroelectrochemistry experiment, the electrolyte of the cell is emptied and filled with 0.01M Ferrocene in 0.1M TBAClO$_4$ in acetonitrile solution. A cyclic voltammogram of ferrocene was taken for reference.

**PL measurement:**

The samples were excited with an 808 nm laser diode modulated by a sine wave at 45 kHz. The signal was collimated by a gold-coated f/1 parabolic mirror and sent through a step-scan Michelson interferometer to a cooled HgCdTe detector. A silicon wafer was placed in front of the detector to filter out the laser. The detector output was sent through a lock-in amplifier. The interferometer was controlled by a step motor and scanned to give spectra with 50 cm$^{-1}$ resolution after Fourier transformation.

**Electrochemical mobility measurements:**

There are four electrodes in the setup: two interdigitated Pt working electrodes, one reference electrode, and one Pt counter electrode. The bipotentiostat (DY2300 series Digi-Ivy) applies a small bias, $V$, (-0.1 V) to the working electrodes and measures the currents of the two working electrodes ($I_1$, $I_2$) as a function of potential. ($I_1 + I_2$) is the charging current of the film while ($I_1 - I_2$)/2 is the conduction current. The conductance (G) is $G = (I_1 - I_2)/2V$. The scanning rate is 50 mV/s. The temperature is monitored by a chromel–alumel thermocouple. An Ag-wire reference electrode is used in the cell. The sample is immersed in an electrochemical cell filled with 0.1M TBAClO$_4$ in acetonitrile and cooled in an ethanol/dry ice bath. As in prior electrochemical studies, cooling is used to minimize the Faradaic current due to impurities. The electrochemical cell was assembled inside a nitrogen-filled glovebox.

**Acknowledgement:**

The work was supported by the grant DE-SC0023210 funded by the U.S. Department of Energy, Office of Science. S.K.S would like to acknowledge the Department of Chemistry, the University of Chicago, for the Martha Ann and Joseph A. Chenicek/ John C. Light Memorial Fellowship. S.K.S. acknowledges Dr. Ananth Kamath, Dr. Jun Hyuk Chang, Mr. Zirui Zhou, and Ms. Tanya Chen for useful discussions and valuable suggestions, and Dr. Xingyu Shen for help with the UV-Vis-NIR absorption measurements.

**Conflict of Interest:**

The authors declare no competing financial interest.

Supporting Information

# Mid-Infrared Intraband transition in InAs colloidal quantum dots

*Shraman Kumar Saha[1], Philippe Guyot-Sionnest[1] **

[1]*Department of Chemistry, The University of Chicago, Chicago, IL, 60653*

Index

**Discussion 01: Fitting of the steady-state Visible-NIR absorption spectra**

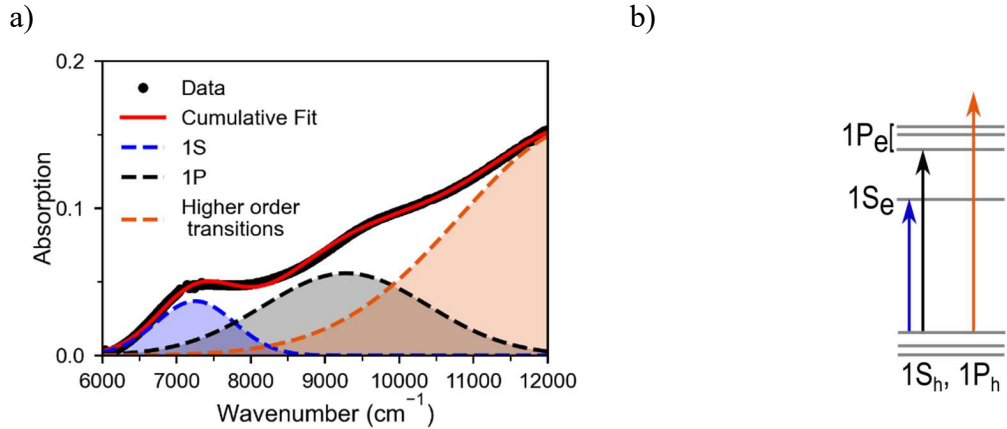

**Figure S01:** a) Peak fitting of the absorption spectra of InAs/InP core/shell dots with three Gaussians. The reduced chi^2 values are in the order of $10^{-6}$. b) The schematic shows the three different transitions accounted for in the fitting.

| Transition | Amplitude (a.u.) | Peak (cm$^{-1}$) | Standard deviation (cm$^{-1}$) |
|---|---|---|---|
| 1S | 0.037 +/- 0.004 | 7250 +/- 30 | 525 +/- 25 |
| 1P | 0.06 +/- 0.05 | 9200 +/- 250 | 1100 +/- 300 |
| Higher order | 0.15 +/- 0.02 | 12500 +/- 500 | 1700 +/- 1000 |

**Table S1:** Fitting parameters for the absorption spectra of InAs/InP dots.

**Discussion 02: Fitting p-XRD peak with pseudo-Voigt profile**

The diffraction peak profile is often described as a combination of Gaussian and Lorentzian functions. A pseudo-Voigt function is a linear combination of these two functions.

The functions are as follows:

Gaussian function: G(x)

$$G(x) \sim e^{\left(-\frac{(x-\mu)^2}{2\sigma^2}\right)}$$

Lorentzian function: L(x)

$$L(x) \sim \frac{1}{\left(1 + \left(\frac{(x-\mu)}{\gamma}\right)^2\right)}$$

Pseudo-Voigt function: F(x)

$$F(x) = A(nL(x) + (1-n)G(x))$$

where, μ is the center of the peak, σ is the Gaussian broadening, γ is the Lorentzian broadening, $n$ is the mixing parameter ($0<n<1$), and $A$ is the peak height.

Debye-Scherrer sizes are calculated using the following formula:

$$D = \frac{0.94\,\lambda}{\beta \cos(\theta)}$$

where λ is the X-ray wavelength (0.154 nm), β is the FWHM of the peak in radians, and θ is the diffraction angle.

| Material | FWHM | Diffraction angle (°) | Debye-Scherrer size | Thickness of shell | Shell thickness |
|---|---|---|---|---|---|
| InAs core | 1.563° → 0.027 rad | 25.506/2 → 12.753 | 5.45 nm | x | X |
| InAs/InP core/shell | 1.405° → 0.024 rad | 25.543/2 → 12.7715 | 6.06 nm | 0.34 nm | 1.8 ML |
| InAs/ZnSe core/shell | 1.251° → 0.022 rad | 25.554/2 → 12.777 | 6.81 nm | 0.33 nm | 2.1 ML |

**Table S2A:** Analysis for Debye-Scherrer size and estimation of shell thickness

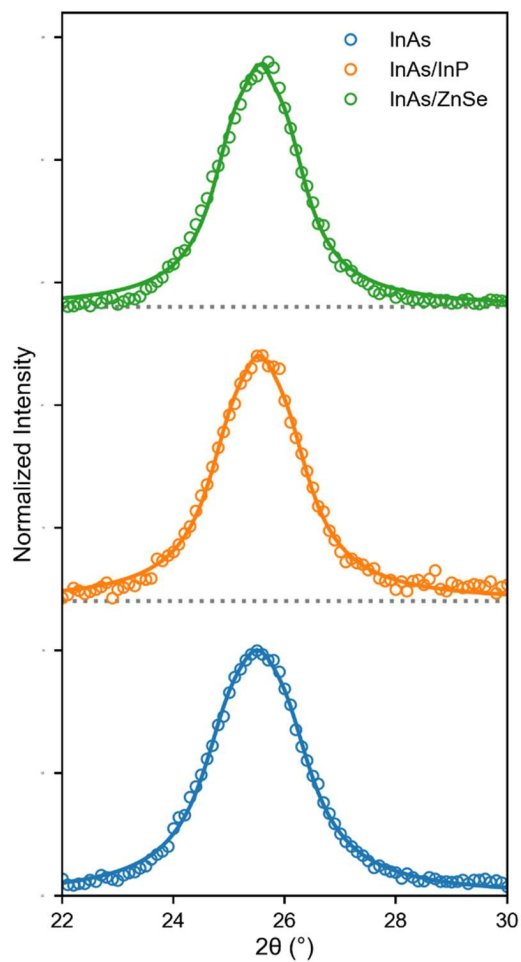

**Figure S02:** Fitting of the pxrd peak with pseudo-Voigt profile for InAs (blue), InAs/InP (red), and InAs/ZnSe (green) dots. The points represent the experimental data, and the line represents the fit. The black dotted lines represent the respective zero lines.

Vegard's law analysis:

| Material | Peak center | InAs lattice parameter | Lattice constant of shell material | Shell thickness |
|---|---|---|---|---|
| InAs core | 25.506/2 | 0.60583 nm | x | |
| InAs/InP core/shell | 25.597/2 | | 0.58687 nm | 0.67 ML |
| InAs/ZnSe core/shell | 25.554/2 | | 0.5667 nm | 0.21 ML |

**Table S2B:** Estimation of shell thickness from Vegard's law

An example calculation:

$$x_{InAs}\, a_{InAs} + (1 - x_{InAs})\, a_{InP} = a_{InAs/InP}$$

where $x_A$ is the mole fraction of $A$, and $a_A$ is the lattice parameter of $A$.

From pXRD data: $x_{InAs}\, 0.6058 + (1 - x_{InAs})0.5869 = 0.5958$

$$x_{InAs} = 47\%,\ x_{InP} = 52\%$$

Let r be the radius of the core and t be the thickness of the shell. From TEM and pXRD, we know the radius of core/shell dots (r + t) and core dots r.

Now,

$$\frac{(r+t)^3 - r^3}{r^3} = \frac{x_{InP}\, \frac{M_{InP}}{d_{InP}}}{x_{InAs}\, \frac{M_{InAs}}{d_{InAs}}}$$

where $M_A$ is the molar mass of $A$, and $d_A$ is the density of A.

Given r = 3.75 nm for InAs core, (r+t) = 3.95 nm: t = 0.2 nm, the thickness of each InP layer is 0.33 nm, the thickness of the InP shell is 0.6 ML. Based on these calculations for 1 ML of InP, the new angle would be 25.67°.

Similarly, for ZnSe: $x_{InAs} = 95\,\%$, $x_{ZnSe} = 5\%$, this gives (r+t) = 3.8, t = 0.05 nm, thickness of each ZnSe layer is 0.34 nm, thickness of the ZnSe shell is 0.17 ML. For 1 ML of ZnSe, the new angle would be 25.9 °.

**Discussion 03: Transmission electron microscopy (TEM) images and size distribution of core and core/shell dots:**

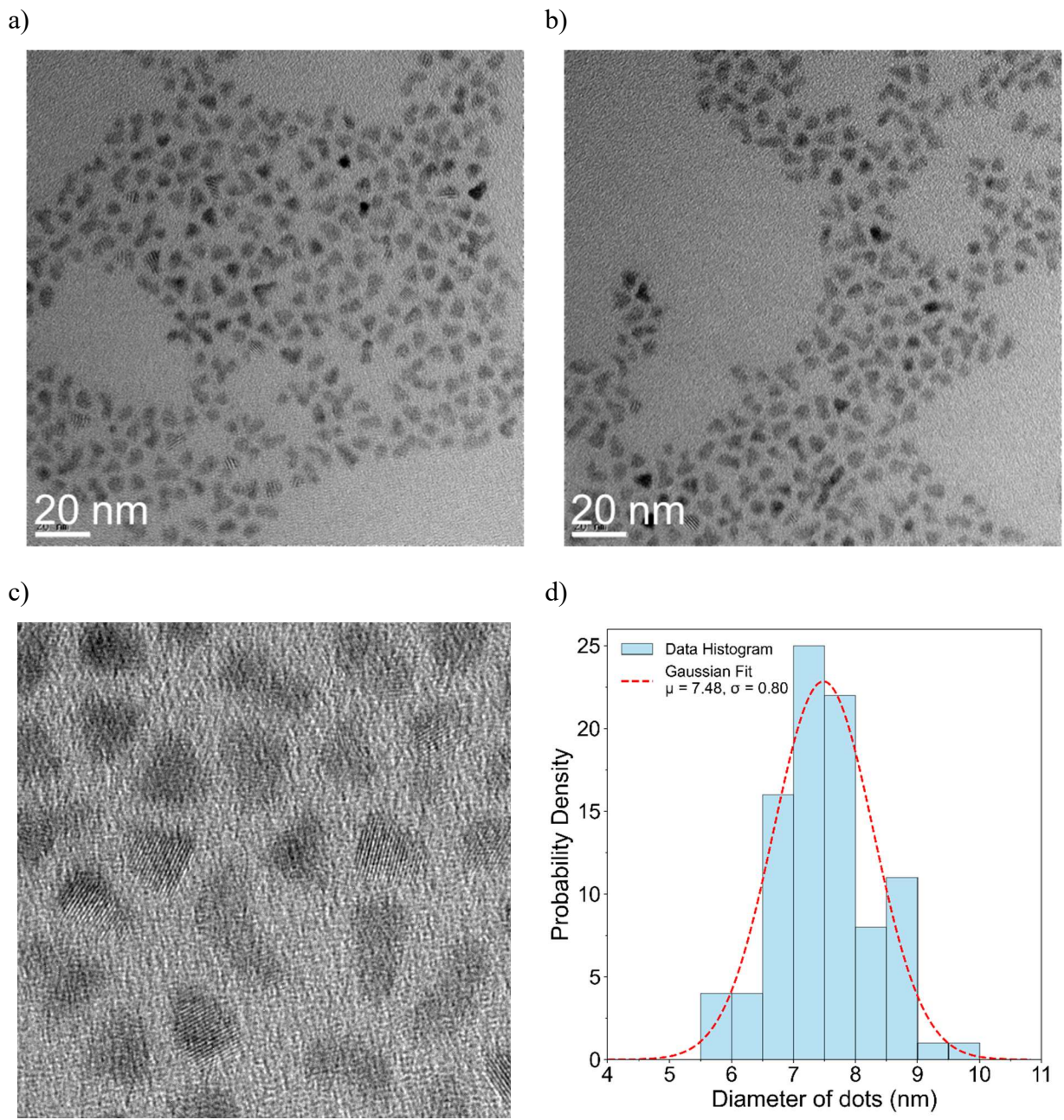

**Figure S03:** a) and b) TEM images of InAs dots at different locations of the TEM grid. c) Magnified individual QDs showing good crystallinity. d) Size distribution histogram of InAs dots obtained from TEM images

a) 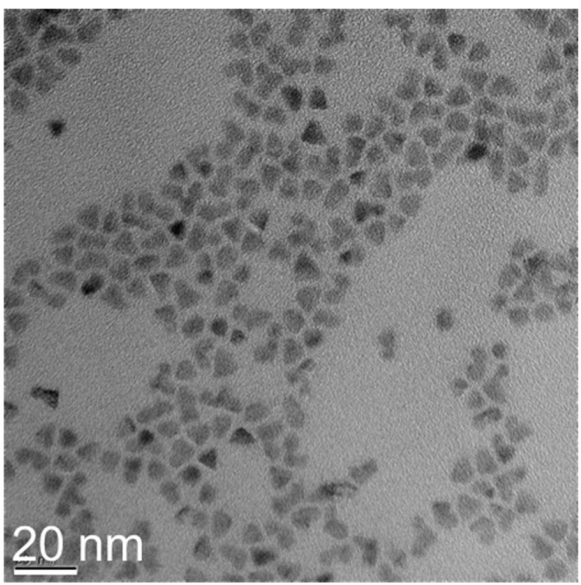

b) 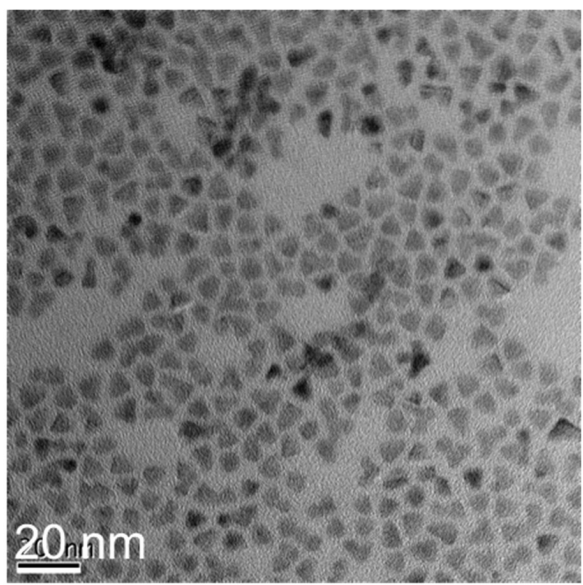

c) 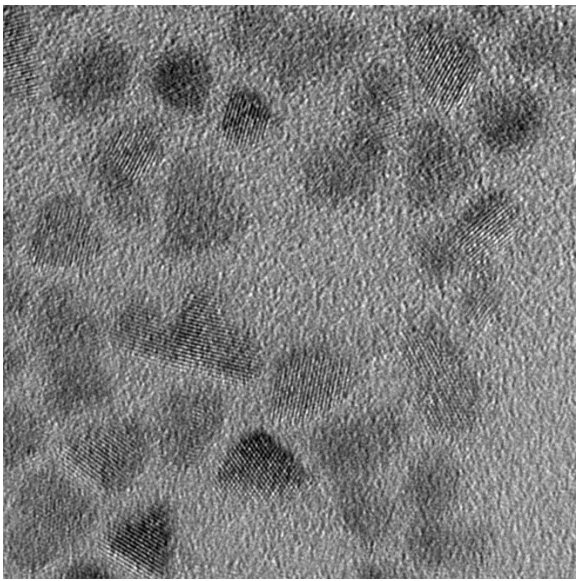

d) 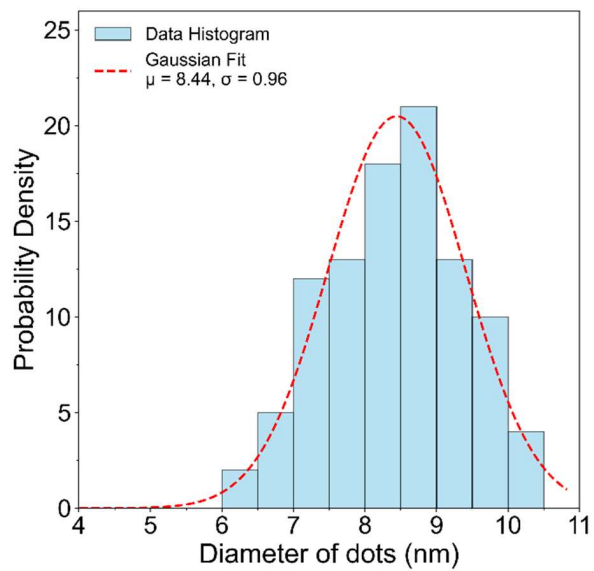

**Figure S04:** a) and b) TEM images of InAs/InP dots at different locations of the TEM grid. c) Magnified individual QDs show good crystallinity. d) Size distribution histogram of InAs/InP dots obtained from TEM images.

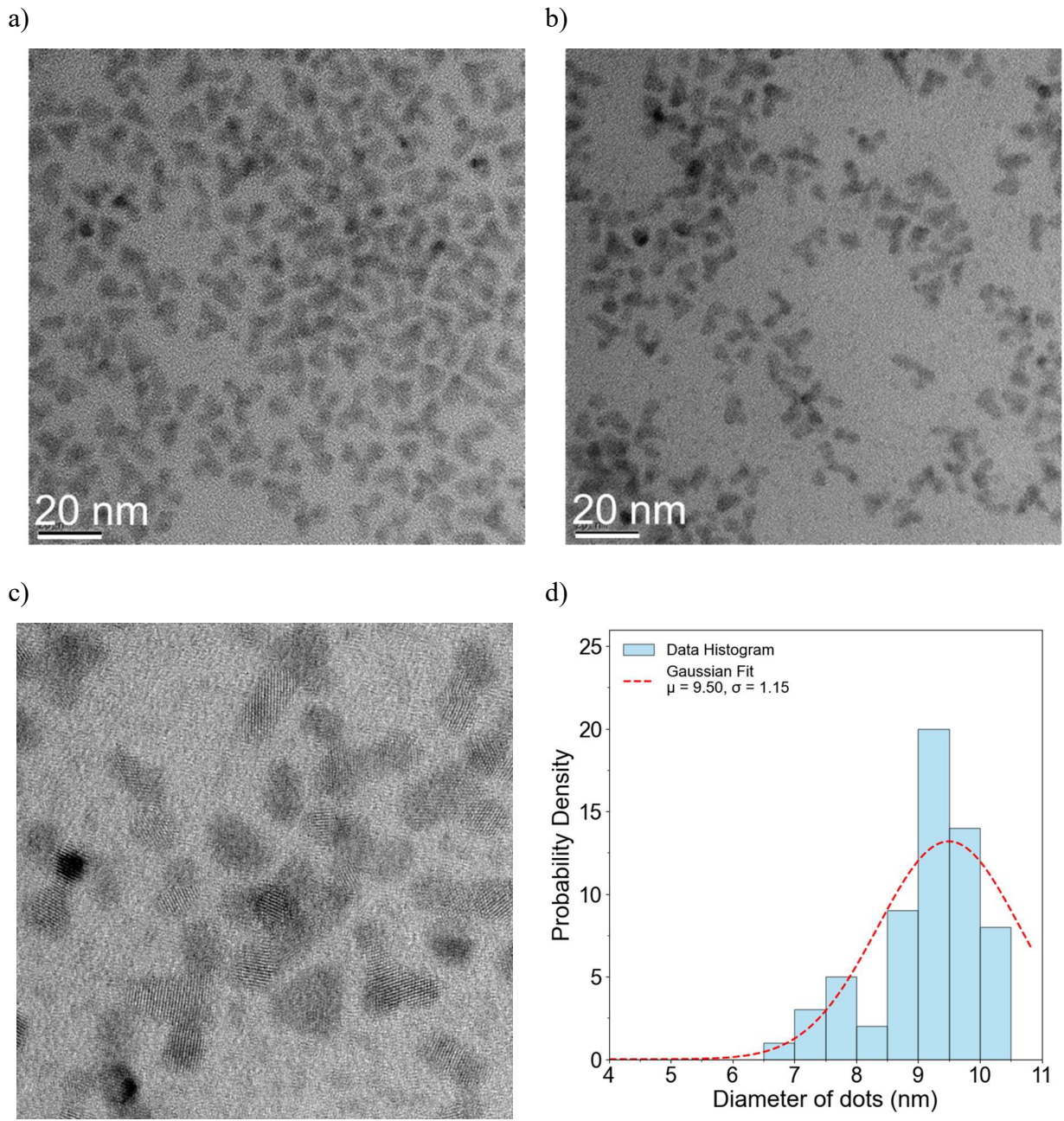

**Figure S05:** a) and b) TEM images of InAs/ZnSe dots at different locations of the TEM grid. c) The magnified image shows good crystallinity. d) Size distribution histogram of InAs/ZnSe dots obtained from TEM images.

| Shell material | Size of core/shell Dots (nm) | Size of InAs dots (nm) | Monolayer (ML) thickness (nm) | Shell thickness (ML) |
|---|---|---|---|---|
| InP | 8.4 +/- 1 | 7.5 +/- 1 | 0.34 | ~ 1.3 |
| ZnSe | 9.5 +/1.15 | | 0.33 | ~ 3 |

**Table S3:** Shell thickness estimation from the sizes obtained from TEM images

**Discussion 04: Calibration of the Silver wire pseudo-reference electrode**

The pseudo reference electrode (PRE) used for the electrochemical experiments is calibrated in 0.1 M TBAClO$_4$ solution in acetonitrile containing 0.01 M ferrocene. The voltage weep rate is 100 mV/s. The cyclic voltammogram of the ferrocene is shown in Figure S6. The $E_{1/2}$ of the ferrocene/ferrocenium ($Fc^+/Fc$) couple is 0.524V vs Ag-PRE.

The scaling of the applied potential (E$_{app}$) with respect to the vacuum level is done by using the following formula:

$$E_{vacuum} = -[\,E_{PRE}(E_{app}) - E_{PRE}(Fc^+/Fc) + E_{NHE}(Fc^+/Fc) + E_{vacuum}(NHE)]$$

The absolute potential of the Normal Hydrogen electrode (NHE) is 4.44V with respect to vacuum.[1] The $E_{1/2}$ of the $Fc^+/Fc$ couple is assumed to be 0.65 V. The measure $E_{1/2}$ of the $Fc^+/Fc$ couple wrt PRE is 0.524V. Substituting these values in the above equation results in:

$$E_{vacuum} = -[\,E_{PRE}(E_{app}) + 4.5]$$

Similarly, the applied bias with respect to NHE will be:

$$E_{NHE} = -[\,E_{PRE}(E_{app}) + 0.13]$$

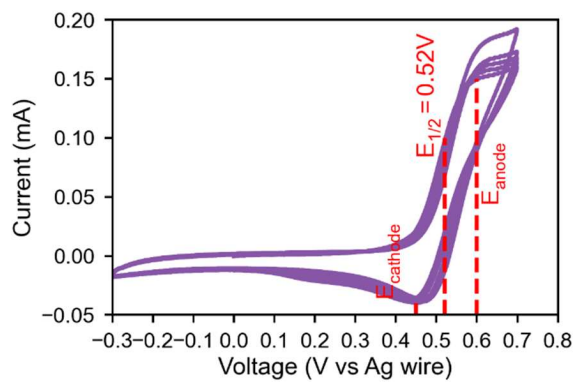

**Figure S06:** Cyclic voltammogram of Ferrocenium$^+$ / Ferrocene in the spectroelectrochemical cell. The $E_{1/2}$ is used to calibrate the Ag pseudo-reference electrode used during the experiment.

**Discussion 05: Gerischer Analysis: The stability of semiconductors under reducing bias**

Following Gerischer, the redox decomposition of semiconductors $MX$ is estimated using the following reaction:

$$(MX)_n + ze^- \rightarrow M^0 + X^{z-} + (MX)_{n-1}$$

Given the Gibbs free energy of formation of $MX$ is $\Delta G_{MX}$ and the standard reduction potential of the counter anion $X$ is $E_{X/X^{z-}} : zH^+ + X^0 + ze^- \rightleftharpoons XH_z$ in aqueous solution, the redox decomposition limit is

$$E_{red} = E_{X/X^{z-}} + \frac{\Delta G_{MX}}{zF}$$

Where $F$ is the Faraday constant, and $z$ is the number of electrons that participate in the reduction process. The redox limit for ZnSe, InAs, and InP is as follows:

| MX | $E_{X/X^{z-}}$ (V vs NHE) | $\Delta G_{MX}$ (KJ/mol) | $\frac{\Delta G_{MX}}{zF}$ (V) | $E_{red}$ (V vs NHE) | $E_{red}$ (eV) |
|---|---|---|---|---|---|
| ZnSe | -0.11 | -177 | -0.91 | -1.04 | -3.4 |
| InP | -0.06 | -77 | -0.27 | -0.33 | -4.1 |
| InAs | -0.23 | -54 | -0.18 | -0.41 | -4.0 |

Table S4: Estimation of decomposition potential for reduction according to the Gerischer model.

These potentials are for aqueous solutions. If the reaction medium can stabilize the free anions, the reduction potential will be lower. The relative stability against reduction is similar for InAs and InP, whereas ZnSe is stable against higher reducing potentials.

**Discussion 06: Sigmodal fitting parameters of Nernst plots**

| Induced Transitions | InAs | | InAs/InP | | InAs/ZnSe | |
|---|---|---|---|---|---|---|
| | Mean potential (eV) | Width (meV) | Mean potential (eV) | Width (meV) | Mean potential (eV) | Width (meV) |
| $i_1$ | -4.17 +/- 0.04 | 110 +/- 30 | -4.21 +/- 0.03 | 120 +/- 30 | -3.97 +/- 0.02 | 160 +/- 20 |
| $i_2$ | -4.18 +/- 0.06 | 100 +/- 30 | -4.22 +/- 0.03 | 120 +/- 30 | -3.92 +/- 0.02 | 170 +/- 20 |
| $e_1$ | -4.26 +/- 0.01 | 77 +/- 10 | -4.32 +/- 0.01 | 60 +/- 04 | -3.98 +/- 0.02 | 120 +/- 20 |
| $e_2$ | x | x | -4.07 +/- 0.03 | 100 +/- 20 | x | x |

Table S5: Sigmoidal fitting parameters for the Nernst plot shown in Figure 2d, e, and f.

# Discussion 07: Δ (ΔOD) spectra

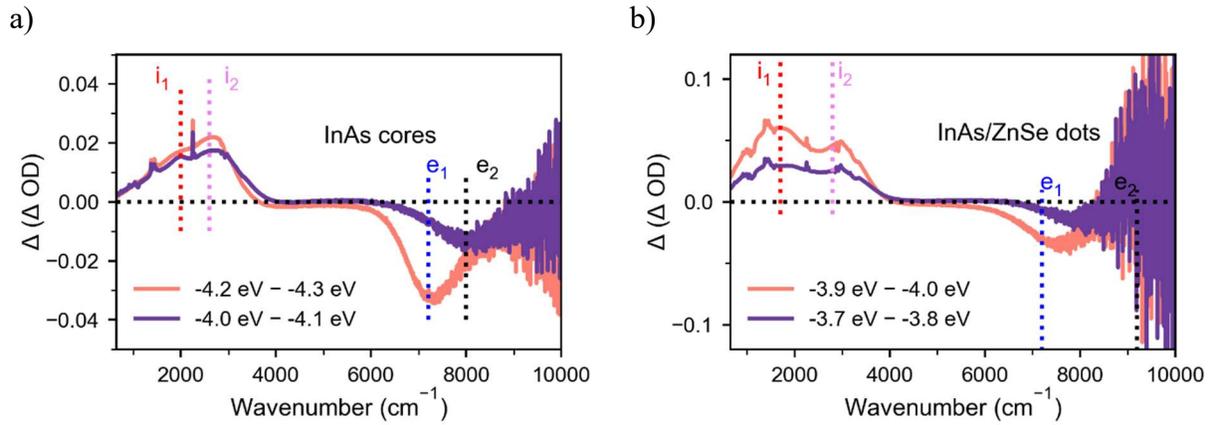

**Figure S07:** Δ (ΔOD) spectra of InAs cores (a) and InAs/ZnSe core/shell dots (b)

# Discussion 08: Two-band k·p model:[2,3]

Solving a two-band **k·p** model yields the following dispersion for the conduction band:

$$E_{CB}(k) = \frac{1}{2}\left(\frac{\hbar^2 k^2}{m_o} - E_G\right) + \sqrt{\frac{\hbar^2 k^2}{2m_o} E_p - \frac{E_G^2}{4}}$$

Where $k$ is the wave vector, $m_o$ is the mass of free electron, $E_G$ is the band gap of the bulk material, $E_p$ is the Kane parameter depends on the effective mass of the electron. $E_p$ is defined as $E_p = \left(\frac{1}{m_e^*} - 1\right) E_G$. From the solution of particle in a sphere of radius r, the $1S_e$ state has $k_{1S} = \pi/r$, $1P_e$ state has $k_{1P} = 4.49/r$, and $1D_e$ state has $k_{1D} = 5.76/r$ where $r$ is the radius of the dots. The energy given by $E_{CB}(k)$ is with respect to the conduction band minimum (CBM).

The heavy hole is assumed to have a parabolic dispersion based on the effective mass model:

$$E_{1S_h}(r) = \frac{\hbar^2 \pi^2}{m_h^* r^2} - E_G$$

$m_h^*$ is the effective mass of holes.

The intraband transition energy ($1S_e \rightarrow 1P_e$) is defined as $E_{intra}(r) = E_{1P_e}(r) - E_{1S_e}(r)$, and interband energy is defined as $E_{inter} = E_{1S_e}(r) - E_{1S_h}(r)$. This neglects electron-electron interactions. Figure S6a shows the intraband vs interband energy. An interband energy around ~ 2000cm$^{-1}$ can be accessed with interband in the SWIR around ~ 6500cm$^{-1}$.

The bulk material properties of InAs are $E_G = 0.35\ eV$, $m_e^* = 0.023\ m_o$, $m_h^* = 0.41\ m_o$, and the absolute value of CBM is -4.9 eV.[4-6] The reduction mean potential is defined as the absolute position of the 1S$_e$ state,

$$Onset\ potential = (E_{1S_e}(r) - 4.9)\ eV\ vs\ vacuum$$

and these neglect charging energy.

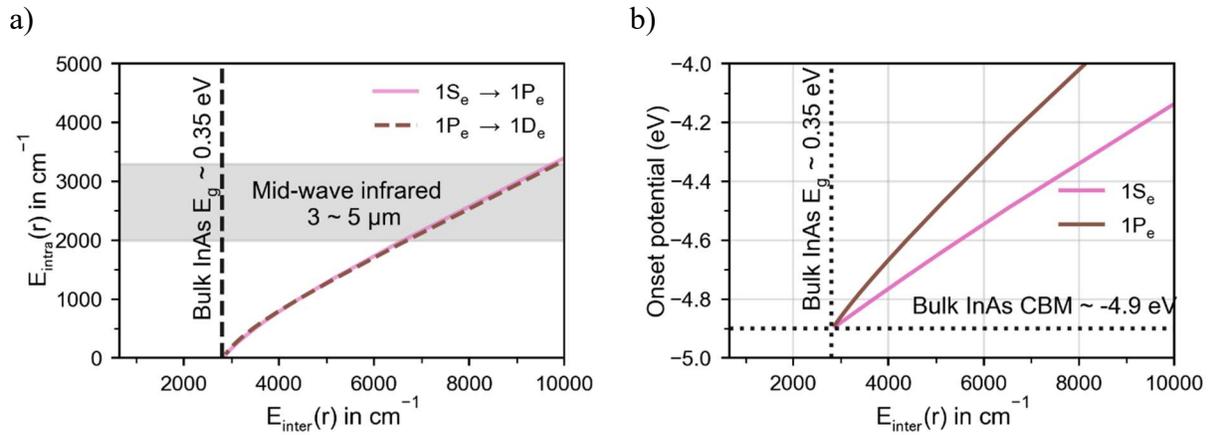

**Figure S08:** a) Intraband energy of InAs CQDs for 1S$_e$ →1P$_e$ (magenta, solid line), and 1P$_e$→1D$_e$ (brown, dashed line). The model predicts that these two intraband transitions have the same energy as a function of the interband transition. b) The mean potential or the absolute position of 1S$_e$ state, and 1P$_e$ state as a function of their interband energy obtained from **k·p** calculations.

**Discussion 09: PL measurements of InAs/InP dots and InAs/ZnSe dots**

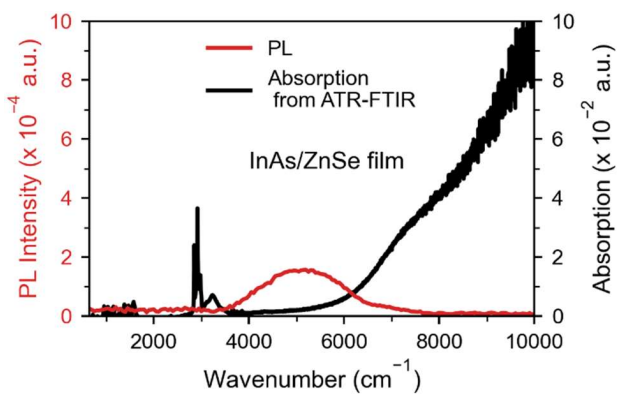

**Figure S09:** Interband photoluminescence (PL) spectrum (red) of InAs/ZnSe dots dispersed in TCE in a CaF$_2$ cuvette. The PL is compared with the absorption spectrum of a film of the same dots (black). There is a surprisingly strong Stokes shift.

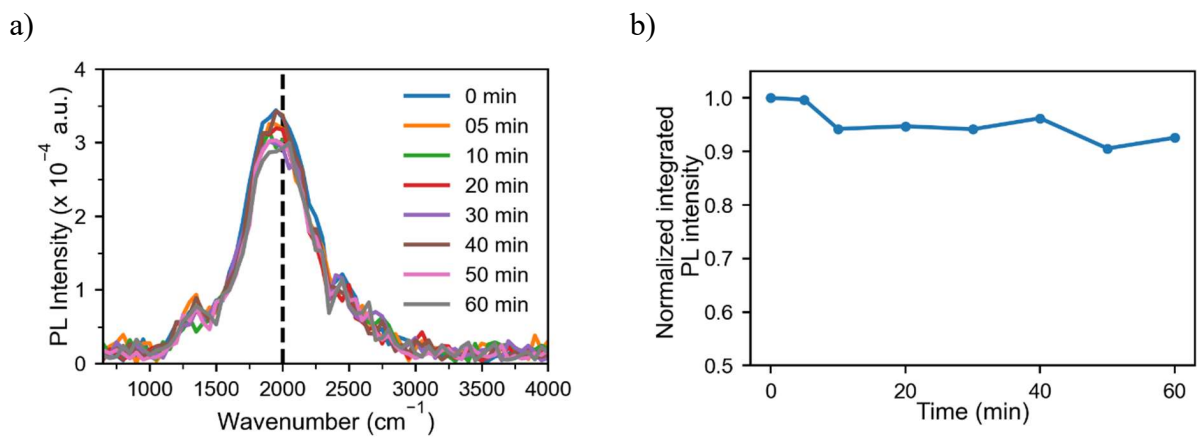

**Figure S10:** a) The intraband PL from the InAs/InP CQDs film upon air exposure. b) The normalized Integrated PL intensity as a function of time.